# THE AUGER EXPERIMENT
# STATUS AND RESULTS

GIORGIO MATTHIAE
on behalf of the Pierre Auger Collaboration
*University and Sezione INFN of Roma Tor Vergata, Roma, Italy*

The Auger experiment was designed to study the high-energy cosmic rays by measuring the properties of the showers produced in the atmosphere. The Southern Auger Observatory has taken data since January 2004. Results on mass composition, energy spectrum and anisotropy of the arrival directions are presented. The most important result is the recent observation of correlations with nearby extragalactic objects.

## 1. Introduction

The flux of cosmic rays, shown in Fig. 1 as a function of energy [1], follows approximately a power law $E^{-\gamma}$ with spectral index $\gamma$ roughly equal to 3.

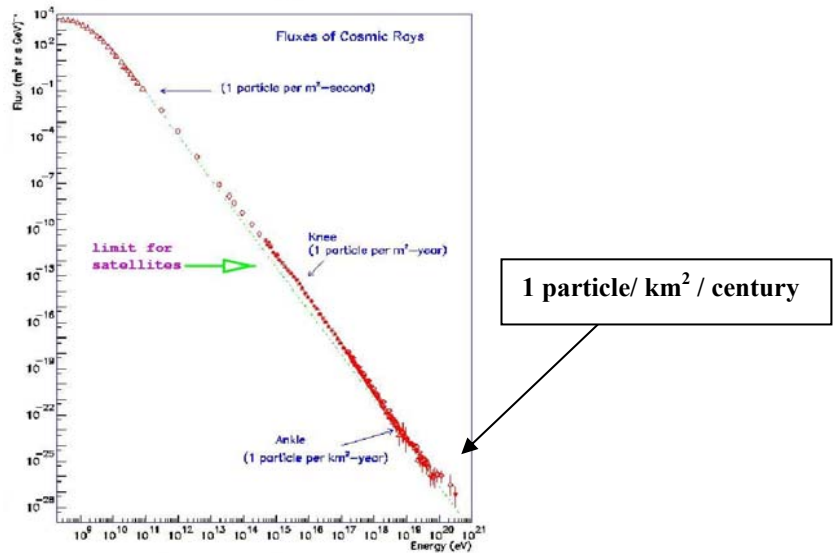

Figure 1. The flux of primary cosmic rays as a function of energy.

The spectrum exhibits interesting features, usually called the "knee" and the "ankle". At the energy of the "knee" ($\sim 3 \times 10^{15}$ eV) the spectral index changes from approximately 2.7 to 3.1. The "ankle", a kind of undulation around a few $10^{18}$ eV, has been actively studied by recent experiments and will be discussed in the Section on the spectrum. In the region above $10^{19}$ eV the flux of the primaries is extremely low, of the order of 1 particle/ km$^2$/ century. Therefore the study of cosmic rays in this very high-energy region requires detectors with very large acceptance.

## 2. The Auger experiment

Two Observatories, one in the Northern and one in the Southern hemisphere are foreseen in the Auger project, to achieve a full exploration of the sky. The Southern Auger Observatory [2] is located near the small town of Malargüe in the province of Mendoza (Argentina) at the latitude of about 35$^0$ S and altitude of 1400 above see level. The region is flat, with very low population density and favorable atmospheric conditions. The Observatory is a hybrid system, a combination of a large surface array and a fluorescence detector.

The surface detector (SD) is a large array of 1600 water Cherenkov counters spaced at a distance of 1.5 km and covering a total area of 3000 km$^2$. Each counter is a plastic tank of cylindrical shape with size 10 m$^2$ x 1.2 m filled with purified water. Technical details of a tank are given in Fig. 2. The surface detector measures the front of the shower as it reaches ground. The tanks activated by the event record the particle density and the time of arrival.

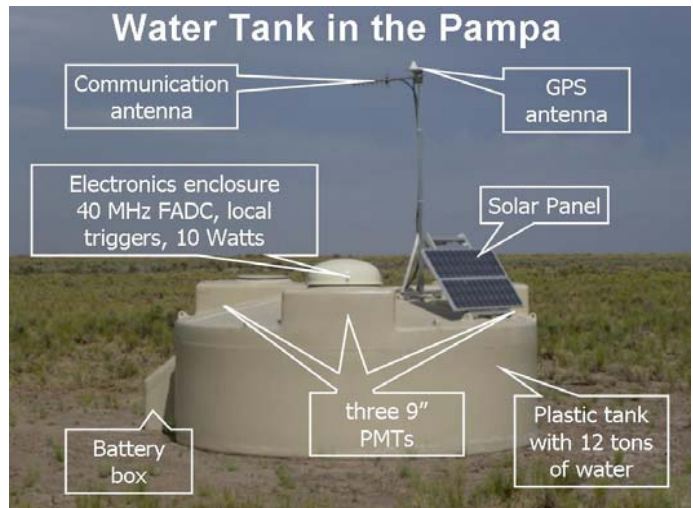

Figure 2. Picture of a water tank of the Surface Detector of the Auger Observatory. The insets give explanations on the various components of the system.

The fluorescence detector (FD) consists of 24 telescopes located in four stations which are built on the top of small elevations on the perimeter of the site. The telescopes measure the shower development in the air by observing the fluorescence light. Each telescope has a 12 m$^2$ spherical mirror with curvature radius of 3.4 m and a camera with 440 photomultipliers. The field of view of each telescope is $30^0$ x $30^0$. UV filters placed on the diaphragm reject light outside the 300-400 nm spectrum of the air fluorescence. The FD may operate only in clear moonless nights and therefore with a duty cycle of about 12%. A sketch of a telescope is shown in Fig. 3.

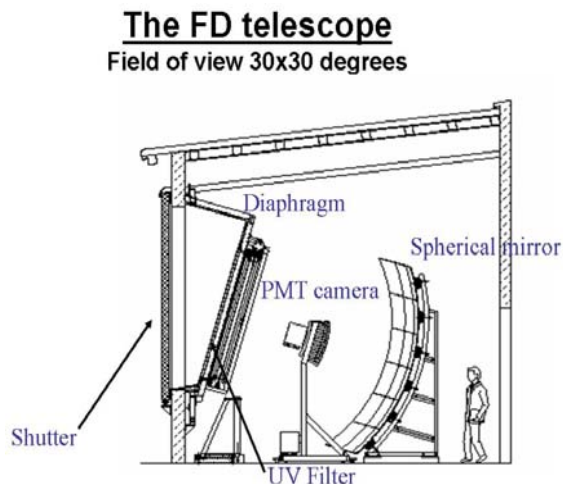

Figure 3. Sketch of a fluorescence telescope. The various components are indicated.

Attenuation of the fluorescence light due to Rayleigh and aerosol scattering along the path from the shower to the telescope is measured systematically with the LIDAR technique.

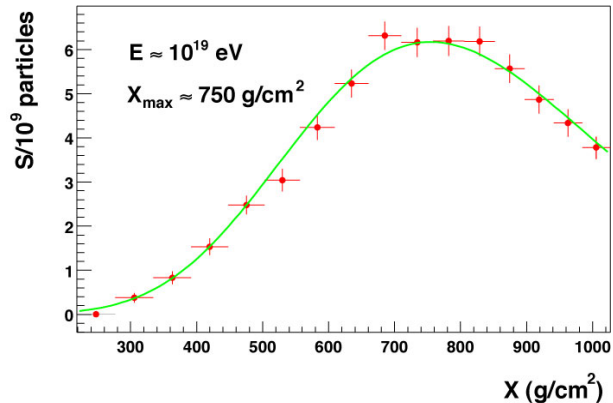

Figure 4. Example of a measured longitudinal profile of a high-energy shower.

An example of a longitudinal profile of a shower as measured by the FD is shown in Fig. 4 where the number of particles of the shower is plotted as a function of the atmospheric depth. In order to obtain the shower profile, the contamination due to Cherenkov light has to be subtracted. The empirical formula by Gaisser and Hillas is used to fit the data.

An example of an event of very high energy as observed by the SD is shown in Fig. 5.

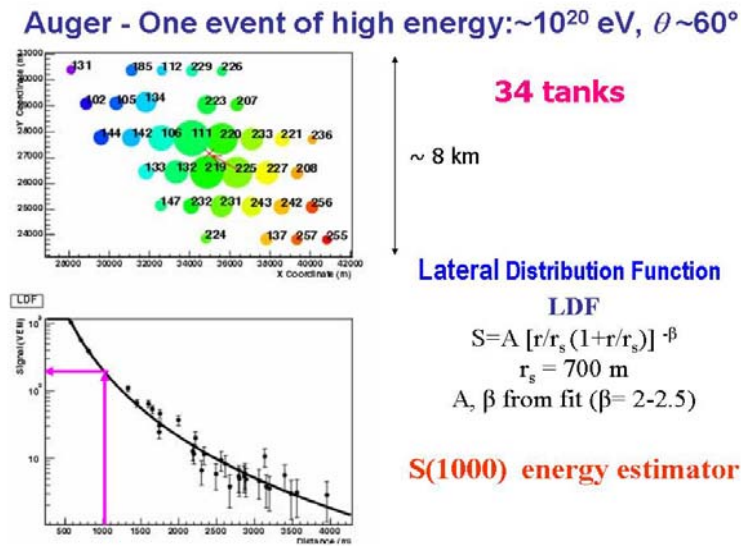

Figure 5. Example of a very high energy event as observed by the SD. The shower has activated 34 tanks distributed over an area of more than 50 km$^2$

The signal of each tank is expressed in units of Vertical Equivalent Muons (WEM) which represents the signal produced by a muon traversing the tank vertically. The flux of cosmic ray muons provides a continuous monitoring of the SD. From the magnitude and the time of the signal of the tanks one derives the direction of the axis of the shower and the point of impact at ground. The left bottom panel of Fig. 5 shows the signal, expressed in units of VEM as a function of the distance from the shower axis. A simple analytical expression known as Lateral Distribution Function (LDF) is then fitted to the data to obtain the signal at the distance of 1000 m from the axis. This interpolated quantity,

S(1000), is a good energy estimator in the sense that it is well correlated with the energy of the primary [3].

The Southern Auger Observatory started to collect data in 2004 and will be completed early in 2008. The Northern Auger Observatory which is now being designed will be located in Colorado (USA).

## 3. Mass composition

The direct method to study the mass composition is based on the measurement of the longitudinal profile of the showers. It is well known that for a given energy protons are more penetrating than light/medium nuclei which interact essentially as a collection of nucleons. The depth of the maximum of the shower profile $X_{max}$, as measured by the fluorescence telescopes, is well correlated with the particle mass. The principle of the method is indicated in Fig. 6. The FD detector of Auger can measure $X_{max}$ with systematic uncertainty of about 15 g/cm$^2$.

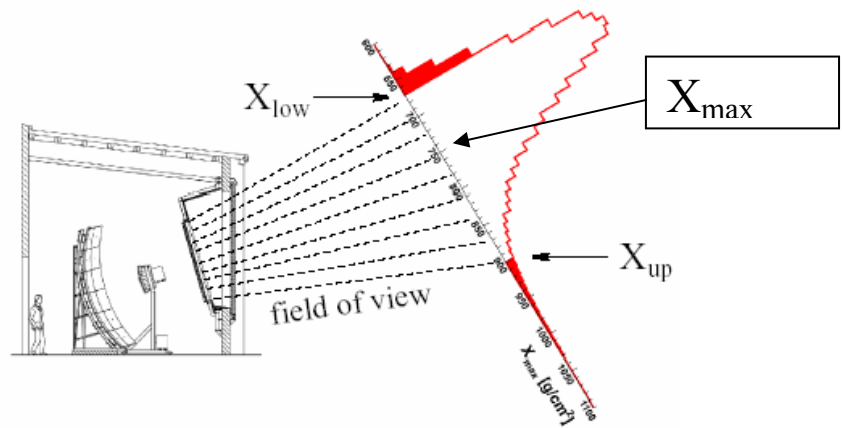

Figure 6. Illustration of the measurement of the quantity $X_{max}$ by a fluorescence telescope of the Auger Observatory .

A compilation of earlier data on $X_{max}$ for energies above $10^{14}$ eV is shown in Fig. 7 where expectations from simulation programs are also given for Fe nuclei, protons and photons. The value of $X_{max}$ for protons is about 100 g cm$^{-2}$ larger than for iron.

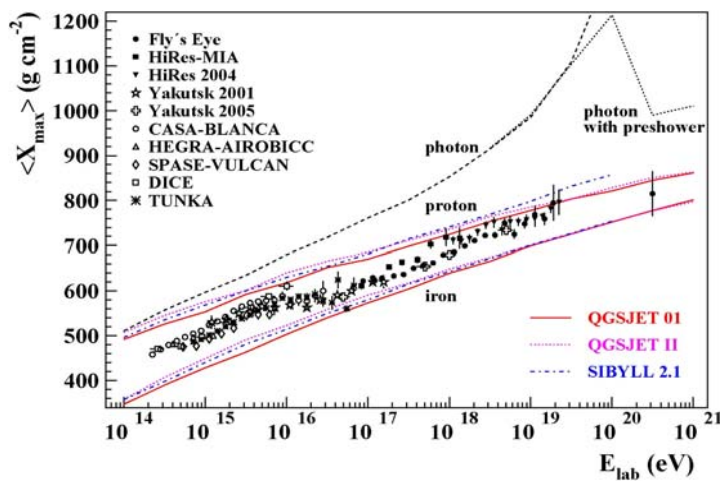

Figure 7. Compilation of earlier data on the quantity $X_{max}$ as a function of energy. Prediction of various simulation programs for incident photons, protons and iron nuclei are also shown.

Recent data by Auger [4] are presented in Fig. 8 together with the predictions of various simulation programs. In spite of the still low statistics, the data indicate some change of regime at 2-3 EeV where the slope (elongation rate) changes. At the highest energies the trend is intermediate between protons and Fe nuclei with a mean mass number of about 5.

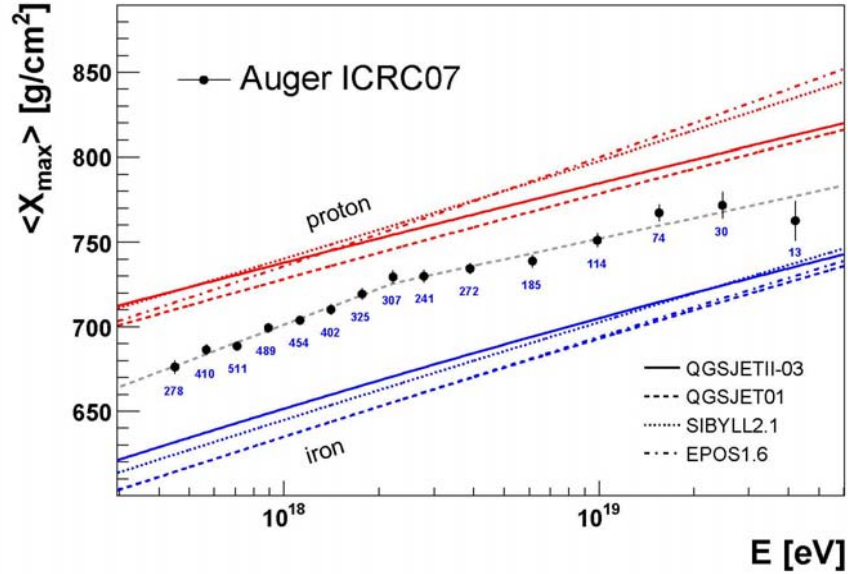

Figure 8. The Auger data on the quantity $X_{max}$ are plotted as a function of energy and compared to predictions of simulation programs for protons and iron nuclei. The number of events for each data point is also shown. The errors shown are statistical.

## 4. The energy spectrum

An important feature of the spectrum in the energy region above $10^{19}$ eV is the mechanism suggested by Greisen, Zatsepin and Kuz'min which is known as GZK cutoff. It is due to the interactions of the cosmic rays with the low energy photons of the Cosmic Microwave Background. Protons with energy above the threshold for photoproduction of pions ($\sim 4 \times 10^{19}$ eV) will lose energy as they travel in space. The value of the energy where an integral power-law spectrum would be reduced to one half is $5.3 \times 10^{19}$ eV [5]. The energy loss per interaction is about 15 – 20 %. At $\sim 5 \times 10^{19}$ eV most of the observed particles must have come from sources within 100 Mpc. Production of electron-positron pairs is also present but it is less effective than photo-pion production. However, this process is expected [5] to be responsible for a feature related to the so-called "ankle", a shallow minimum (or "dip") in the plot of the flux times $E^3$ which is centered at energies of a few $10^{18}$ eV.

In the past there was a controversy on the actual presence of the suppression due to the GZK cutoff. The AGASA data did not show a suppression, contrary to the preliminary data of HiRes. The experimental situation is now clarified by the final data of HiRes [6], shown in Fig. 9 and by the data of Auger. The HiRes data clearly show a steepening of the spectrum above $10^{19.6}$ eV with a fitted value of the spectral index $\gamma = 5.1 \pm 0.7$. The steepening agrees with the expectations from the GZK cutoff.

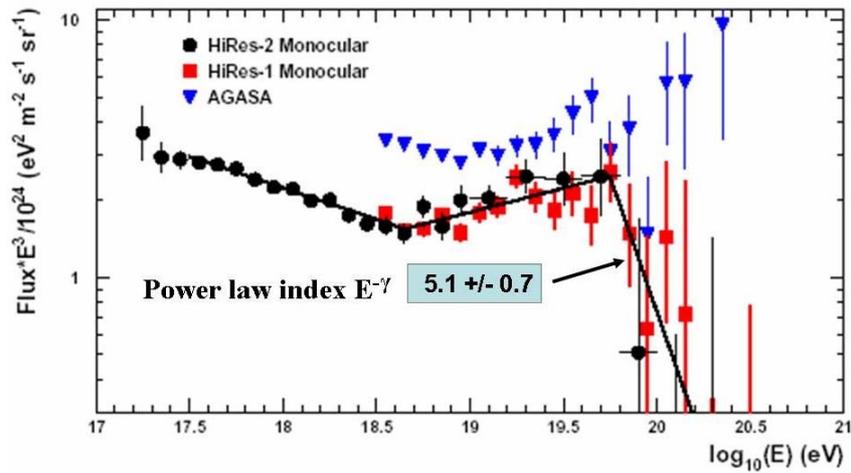

Figure 9. The final HiRes results on the energy spectrum are presented as Flux x $E^3$ and compared to the earlier AGASA data. The steepening due to the GZK cutoff is clearly seen. In addition the shallow minimum centered around $10^{18.6}$ eV is also evident.

The method used by Auger to measure the energy spectrum exploits the hybrid nature of the experiment with the aim of using the data itself rather than simulations.

For each event, the energy estimator S(1000) is obtained as discussed in Section 2. The energy estimator S(1000) depends on the zenith angle because the effective atmosphere thickness seen by showers before reaching ground changes with the zenith angle. The value of S(1000) corresponding to the median zenith angle of $38^0$ is used as reference and the zenith angle dependence of the energy estimator is determined assuming that the arrival directions are isotropically distributed. This procedure is traditionally called "Constant intensity cut".

The absolute calibration of S(1000) is derived from the hybrid events using the calorimetric energy measured by the FD which is then corrected for the missing energy (neutrinos and muons) using the mean value between proton and iron (10% correction at $10^{19}$ eV with uncertainty ± 2%). This absolute calibration, which defines the energy scale, is at present affected by a systematic error of about ± 20%, mainly due to uncertainties on the fluorescence yield and on the calibration of the FD telescopes.

The energy calibration, obtained from the subset of hybrid events (see Fig.10) is then used for the full set of events with higher statistics as measured by the SD.

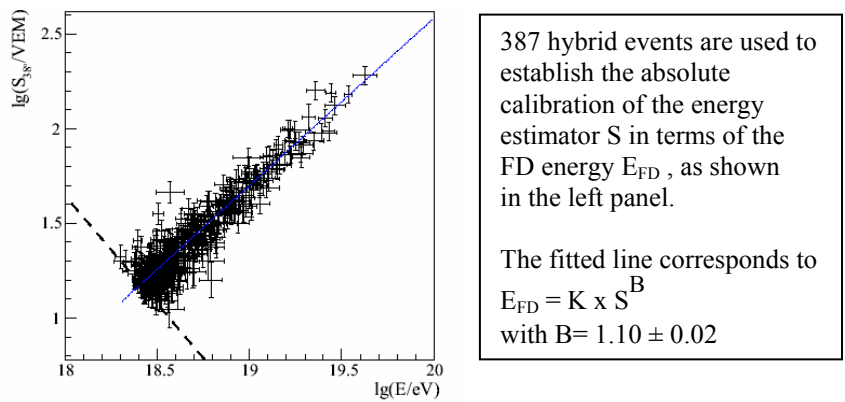

387 hybrid events are used to establish the absolute calibration of the energy estimator S in terms of the FD energy $E_{FD}$, as shown in the left panel.

The fitted line corresponds to
$E_{FD} = K \times S^B$
with B= 1.10 ± 0.02

Figure 10. Calibration of the energy estimator S(1000) using the energy from the FD.

Three different measurements of the energy spectra were obtained by the Auger Collaboration as explained in detail in ref.7. The SD "vertical" spectrum is for zenith angle $\theta < 60^0$, the SD "inclined" is for $60^0 < \theta < 80^0$. The SD spectra starts at the energy of $3 \times 10^{18}$ eV where the efficiency goes to a plateau. The hybrid spectrum refers to FD events with at least one SD tank and starts at a lower energy ($10^{18}$ eV).

The three spectra are consistent within statistics, as shown in Fig. 11 and therefore were combined in a single spectrum which is presented in Fig.12.

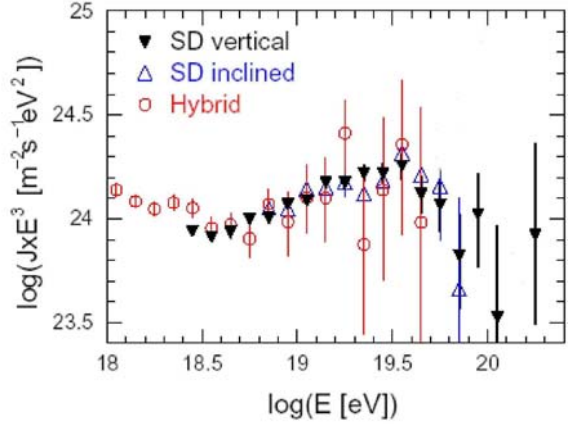

Figure 11. The Auger spectra (flux x $E^3$) from the SD and from hybrid events.

The "ankle", a barely noticeable undulation in Fig.12, appears as a clear shallow minimum in Fig.11 with shape and position similar to the predictions of ref.4. Above $\sim 10^{19.6}$ eV, the steepening expected from the GZK mechanism is also evident.

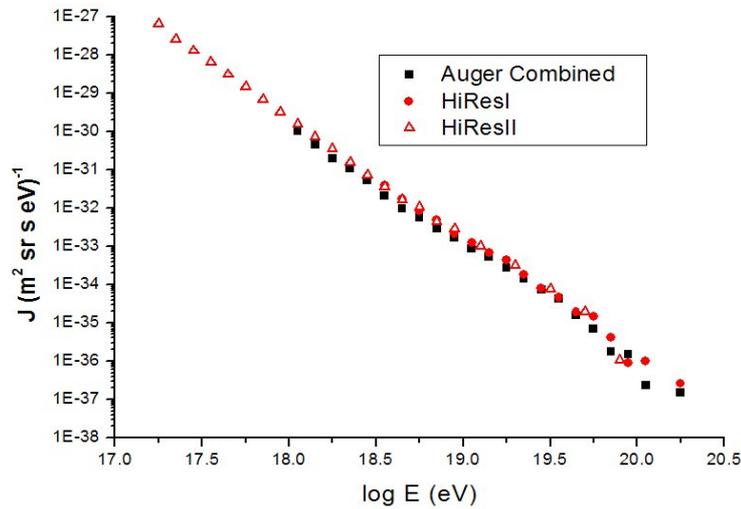

Figure 12. The combined Auger spectrum is plotted together with the HiRes data. The difference between the two sets of data can be attributed to ~10% difference in the energy calibration which in turns is due to different values used for the fluorescence yield.

The structures present in the Auger spectrum are better analyzed by taking the relative difference of the data with respect to the reference form $J_s = A\,E^{-2.6}$. The result is presented in Fig.13. Numerical values of the spectral index $\gamma$ in the different energy intervals are given in Table 1.

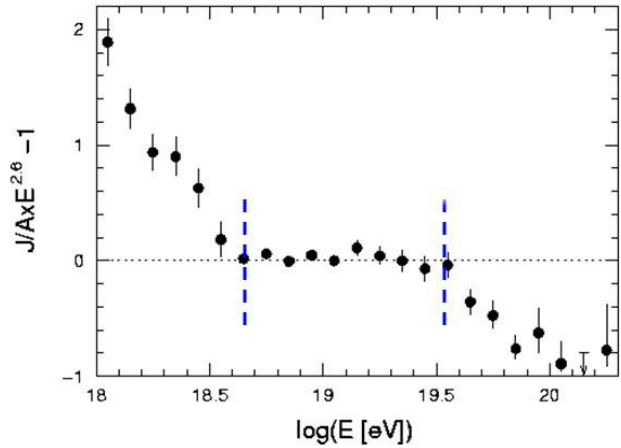

Figure 13. The relative difference of the Auger energy spectrum with respect to the form $E^{-2.6}$. The vertical dashed lines indicate the different intervals for the power law fits.

**Table 1.** Numerical values of the spectral index $\gamma$ of the power law fits in the different energy intervals. The energy values $E_{ankle}$ and $E_{GZK}$ correspond to the position of the breaks.

|  | Auger | HiRes |
|---|---|---|
| $E_{ankle}$ (eV) | $4.5 \times 10^{18}$ | $4.5 \times 10^{18}$ |
| $E_{GZK}$ (eV) | $3.5 \times 10^{19}$ | $5.6 \times 10^{19}$ |
| $\gamma$ ($E < E_{ankle}$) | $3.30 \pm 0.06$ |  |
| $\gamma$ for ($E_{ankle} < E < E_{GZK}$) | $2.62 \pm 0.03$ | $2.81 \pm 0.03$ |
| $\gamma$ for ($E > E_{GZK}$) | $4.1 \pm 0.4$ | $5.1 \pm 0.7$ |

## 5. Anisotropy studies

In the study of anisotropy the Auger Observatory may exploit the good angular resolution of the SD which is better than one degree at high energy.
    Observation of an excess from the region of the Galactic centre at the level of 4.5 $\sigma$, in the energy region 1.0 – 2.5 EeV and with angular scale of $20^0$, was reported by AGASA [8]. The Auger Observatory is suitable for this study because the Galactic centre (constellation of Sagittarius), lies well in the field of view of the experiment. Some of the Auger results [9] on the observed and the expected number of events in the direction of the Galactic centre are shown in Table 2. Clearly the Auger data don't confirm the AGASA result.

**Table 2.** The number of observed and of expected events and the corresponding ratios are listed for different angular windows in the direction of the Galactic centre (energy between 1 and 10 EeV)

| Angular window (degrees) | $N_{observed}$ / $N_{expected}$ | Ratio (errors: stat, syst) |
|---|---|---|
| 5 | 425/393 | $1.08 \pm 0.07 \pm 0.01$ |
| 10 | 1662/1578 | $1.05 \pm 0.04 \pm 0.01$ |
| 20 | 6365/6252 | $1.02 \pm 0.02 \pm 0.01$ |

    The Auger collaboration has done an extensive search for correlation of the high-energy events with known astrophysical objects. This study started early in 2004 and the results from data collected until August 2007 have been published recently [10]. During this period, the Observatory has increased in size. The total exposure is about 20% larger than for the yearly exposure of the Observatory once completed.
    The simple plot of Fig.14 already gives a hint that high-energy events are not distributed isotropically but rather tend to concentrate on the supergalactic plane where most of the nearby galaxies are located.

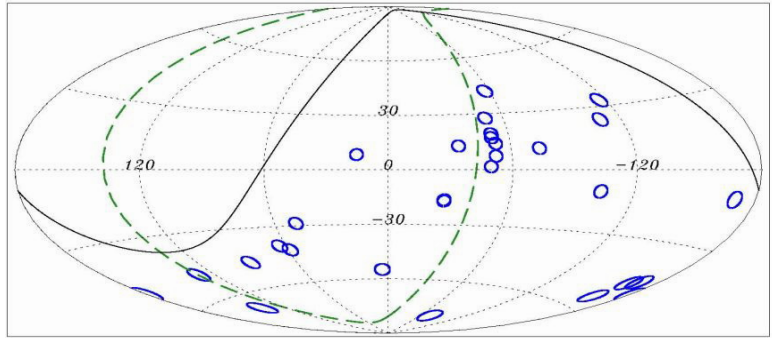

Figure 14. Plot in galactic coordinates showing the high-energy events as small blue circles. The supergalactic plane is indicated by the green dashed line.

A more complete picture showing the data and the position of nearby AGN (from the catalog of ref.11) is reported in Fig.15. Two events are correlated within less than 3 degrees with Cen A, a strong radio source at the distance of about 4 Mpc.

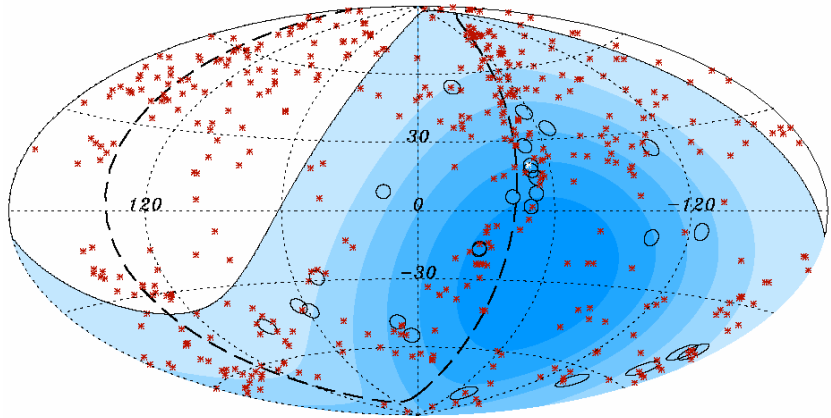

Figure 15. Plot in galactic coordinates showing the events with energy larger than 57 EeV as small circles of radius 3.2 degrees. The supergalactic plane is shown as a dashed line. The red crosses indicate the position of AGN within 71 Mpc. Cen A, one of the nearest AGN is marked in white. The white region of the sky is not accessible from the Southern Auger Observatory. Darker blue regions indicate larger relative exposure.

A sophisticated analysis described in ref. 10 has shown that a clear correlation, within an angle $\psi$ about equal to 3 degrees, exists between the arrival directions of cosmic rays with energy above about 60 EeV and galaxies with active nuclei (AGN) at distances less than about 75 Mpc.

The results are summarized in Table 3. The first exploratory analysis has shown that 12 out of 15 events with energy above 57 EeV were correlated with AGN at distances less than 75 Mpc, within 3.1 degrees while only 3.2 were expected to be correlated by chance for an isotropic distribution.

As a consequence of this result, a prescribed test was defined to see whether the isotropy hypothesis had to be accepted or rejected. The same set of parameters and the same reconstruction algorithms were used. The second independent set (see Table 3, row #2) satisfied the test and the probability for this single configuration to happen by chance if the flux was isotropic is $1.7 \times 10^{-3}$.

A complete reanalysis of the data set gave the results reported in Table 3, row#3. Out of 27 events, 20 were found to correlate with a chance probability of the order of $10^{-5}$.

The correlation becomes statistically more significant if the events in the region around the galactic plane ($|b| < 12$ degrees) are removed. For this subset of 21 events, 19 are correlated with AGN. Elimination of the galactic plane region is motivated by the incompleteness of the catalog in this region and by the expected stronger effect of the galactic magnetic field which is known to be concentrated in the galactic disk.

Table 3. Results of the analysis for the first set, the second independent set, the reanalysis of the full set and for the full data set excluding the galactic plane region are reported.

|  | Number of events E >57 EeV | Events correlated with AGN $\psi$ = 3.1 degree | Events expected for isotropy |
|---|---|---|---|
| Exploratory scan 1 Jan 04- 27 May 06 | 15 | 12 | 3.2 |
| Second independent set 27 May 06–31 Aug 07 | 13 | 8 | 2.7 |
| Full data set (about 1.2 year full Auger) | 27 | 20 | 5.6 |
| Full data set excluding galactic plane region | 21 | 19 | 5.0 |

The distribution of the separation angle between the direction of the 27 high-energy events and the nearest AGN is shown in Fig. 16. The histogram of the data shows a clear departure from isotropy.

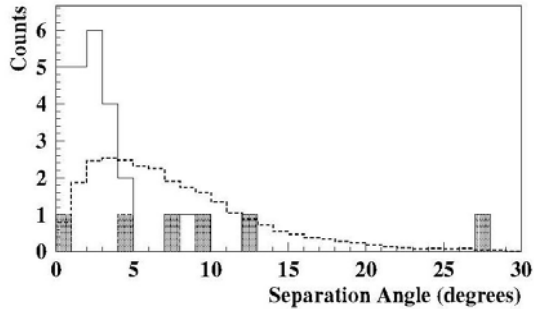

Figure 16. The angle between each event and the nearest AGN. The dotted histogram represents the expectation for isotropic distribution. The histogram shows the data while the 6 shaded areas represent the events removed because close to the galactic plane.